\documentclass[pre,showpacs,twocolumn]{revtex4}

\usepackage[dvips]{graphicx}
\usepackage{amsmath,psfrag}

\usepackage[ps2pdf]{hyperref} 
\hypersetup{colorlinks=false, pdfpagemode=UseOutlines,
 bookmarksopen=false,          bookmarksnumbered=true, pdfstartpage= {1},
 pdftitle    = {Attractor length distribution of critical Boolean networks},
 pdfsubject  = {Work in process},
 pdfauthor   = {florian.greil<at>physik.tu-darmstadt.de},
 pdfkeywords = {Institut fuer Festkoerperphysik, Technische Universitaet Darmstadt, Germany},
 pdfcreator  = {Adobe-Acrobat-Distiller},
 pdfproducer = {LaTeX with hyperref}
}

\renewcommand{\marginpar}[1]{\typeout{#1}}

\newcommand{\lcm}{{\rm LCM}}
\begin{document}
\title{Attractor period distribution for critical Boolean networks}
\author{Florian Greil}
\thanks{\url{florian.greil@physik.tu-darmstadt.de}}
\affiliation{Institut f\"ur Festk\"orperphysik, Technische Universit\"at
Darmstadt, D-64285 Darmstadt, Germany\\
\emph{current address:} Lehrstuhl f\"ur Bioinformatik, Universit\"at Leipzig,
D-04107 Leipzig, Germany}
\author{Kevin E.\ Bassler}
\thanks{\url{bassler@uh.edu}}
\affiliation{Department of Physics, University of Houston, Houston, Texas
77204-5005, USA} 
\affiliation{Texas Center for Superconductivity, University of
Houston, Houston, Texas 77204-5002, USA}
\begin{abstract}
Using analytic arguments, we show that
dynamical attractor periods in large critical Boolean
networks are power-law distributed.  Our arguments are based on the method of
relevant components, which focuses on the behavior of the nodes that control
the dynamics of the entire network and thus determine the attractors.  Assuming
that the attractor period is equal to the least common multiple of the size of
all relevant components, we show that the distribution in large
networks is well approximated by a power-law with an exponent of $-1$.
Numerical evidence based on sampling of attractors supports the conclusions
of our analytic arguments.
\end{abstract}
\maketitle

Boolean networks have been extensively studied as simplified models for complex
systems of multiple interacting units
\cite{albert:statistical,aldana-gonzalez:boolean,drossel:random}. 
They are defined by a directed graph in which the
nodes have binary output states that are determined by Boolean functions of
the states of the nodes connected to them with directed in-links.
They have been used to model a variety of biological, physical, and social
systems.
In a number of recent studies
\cite{li.long.ea:yeast,albert.othmer:topology,davidich.bornholdt:boolean}, for
example, it has been shown that Boolean network models can correctly reproduce
essential features of the dynamics of real biological networks, while dramatically
limiting the number of parameters needed to describe their
behavior \cite{bornholdt:less}.

In their original variant, the out-links of the directed graph that describes
the topology of Boolean networks were assumed to be completely random.  Of
course, the topology of the interactions in real networks is far from being
random.  Nevertheless, the extreme abstraction of random Boolean
networks allows for the identification of basic mathematical laws of
complex network dynamics.  Many non-trivial effects occur in this seemingly
trivial model.  Only recently major progress has been made toward analytically
understanding the dynamics of random Boolean networks
\cite{samuelsson:superpolynomial,drossel:number,mihaljev.drossel:scaling,drossel.mihaljev.ea:number}.

Here we study the dynamical attractor distribution of critical
Boolean networks 
using the method of relevant components. Numerical
evidence is presented that supports our analytic arguments. 
Additionally, we show that the attractor distributions occurring in Boolean
networks that have evolved to criticality based on a competition between
nodes
\cite{paczuski.bassler.ea:self-organized,bassler.lee.ea:evolution,liu.bassler:finite}
match those we can deduce from the analytical understanding how attractors
arise.

Consider ensembles of networks with $N$ nodes in which the behavior
each node depends on exactly $K$ other nodes.  Thus, each node of the directed
graph defining the network interactions has exactly $K$ in-links.  The Boolean
states of the nodes are updated synchronously at uniformly spaced time steps
that can be assumed to be of unit duration.  Each node $i$ has a Boolean state
$s_i(t)\in \{ 0,1\}$ at time $t$ that is determined by a Boolean function $f_i$
of the states of the $K$ nodes it depends on $i_1$, $i_2$, \ldots $i_K$ had
during the previous time step
\begin{eqnarray}
s_i(t+1) = f_i\left( s_{i_1}(t), s_{i_2}(t), \ldots, s_{i_K}(t) \right),
\qquad \forall i.
\end{eqnarray}
The choice of the functions~$f_i$ determines the dynamics. A particular directed
graph together with the Boolean functions defined at each node is
a \emph{network realization}, and so we consider consider ensembles of
network realizations.

The network state of a given realization at time $t$ is the vector of $N$ Boolean
components given by the states of the nodes at that time, $\vec{s}(t)$.  There are $2^N$
possible network states.  The networks we consider, therefore, have a finite
state space and a discrete, deterministic dynamics.
Thus, starting from any initial state the dynamics of the network state
will, in finite time, collapse to an attractor of finite period.
There can be one, or more, attractors, each of which has a basin of attraction
corresponding to the region of state space from which the dynamics 
eventually collapses to that attractor.
The late time dynamics of a Boolean network can be quantified by the number of
attractors $\nu$, their periods $L_j$, and the size of their basins of
attraction $A_j$, $j=1,2,\ldots,\nu$.

Generally, depending on the sets of Boolean functions describing the
dynamics of the nodes ${\cal F}=\{f_1, f_2, \ldots, f_N\}$ in the
realizations of an ensemble of networks, the dynamics of the ensemble can
be classified into two phases with distinct behavior. We want to focus on
the boundary between the phases, where the dynamics is \emph{critical}.  A
critical network can be obtained either by construction, by evolving the
set ${\cal F}$ of Boolean functions
\cite{bassler.lee.ea:evolution,paczuski.bassler.ea:self-organized,liu.bassler:finite}
or by evolving the topology
\cite{bornholdt.rohlf:topological,liu.bassler:emergent,gross.blasius:adaptive,luque.ballesteros.ea:self-organized}.
\marginpar{(R)}
Note though that evolving the Boolean functions can effectively result in
an evolution of the topology 
\cite{reichhardt.bassler:canalization}.

The paper is organized as follows: First, we introduce some needed concepts
and briefly review what is currently known about the attractor distribution
of critical Boolean networks.  These prior results are all numerical.
Second, the attractor period
distribution is calculated using novel analytic methods.  
As we will see, consistent with previous
numerical observations, the attractor periods in large critical Boolean
networks are power-law distributed.  Finally, numerical results, obtained
using novel methods, are shown that confirm our analytic predictions and
allow some of the previous numerical results to be explained.

\typeout{Concepts} 

It is possible to distinguish the two phases of the
dynamics in random Boolean networks by considering the sensitivity of the
dynamical trajectory of the network state to a small change
\cite{aldana-gonzalez:boolean,derrida:random,shmulevich.kauffman:activities}.
Consider two different initial states of a single network realization, 
$\vec{s}^{(1)}(0)$ and $\vec{s}^{(2)}(0)$.
The normalized Hamming distance~$h_t$ between subsequent trajectories of the two
states is the fraction of nodes having a different node value: $h(t) = N^{-1}
\sum_{i=1}^N \left(s_i^{(1)}(t) - s_i^{(2)}(t)\right)^2$.  For small $h(t)$
the probability that more than one input of a node differs in the
two states can be neglected 
and the time evolution of the Hamming distance can be written as $h(t+1) = \lambda h(t)$
where $\lambda$ is the \emph{sensitivity} \cite{shmulevich.kauffman:activities}.

The ``frozen'' phase of the dynamics
is characterized by small sensitivities, $\lambda < 1$.  In this case,
a perturbation of a node's value propagates to less than one other node on
average per time step.  In this phase, in a large network, the output
states of the vast majority of nodes become frozen and a perturbation of a
node's value will eventually die out.
A node is ``frozen'' on an attractor if it stops
changing its value after some transient time.  
The number and periods of the attractors of the network are not influenced by
frozen nodes.  Instead, as we shall see shortly, those quantities can be found
from combinatorial arguments involving only the number of ``relevant''
non-frozen nodes.  As there are few of those in the frozen phase, the
attractors will therefore be short.

On the other hand, in the ``chaotic'' phase the sensitivity
is large, $\lambda > 1$, and a perturbation of a
node's state spreads on average to more than one node each time step. Even
after long times there will still be nodes changing their state because of the
perturbation and so there is a lot of non-frozen nodes.  Thus, attractor
periods can be very long, incorporating a finite fraction of the whole state
space.

However, for networks with $K > 2$, if the functions are chosen with
a bias toward having homogeneous output regardless of their input,
then frozen behavior can also occur. In particular,
as the homogeneity increases, the sensitivity decreases, and
at a critical value of homogeneity a phase transition
from chaotic to frozen behavior occurs
\cite{derrida:random}. Here, we are interested in critical
networks that are at the boundary between the two phases.  As we will see,
critical networks have a rich, complex dynamical behavior that is
intermediate between frozen and chaotic~\cite{aldana-gonzalez:boolean}.

\typeout{> Relevant components.} 
In order to understand the dynamics of a Boolean network it is important to
find the relevant components of the network that control its dynamics.  This
``method of relevant components'' is based on a classification of nodes according
to their dynamical behavior \cite{bastolla:modular}.  There are three kind of
nodes: frozen nodes and two kinds of non-frozen nodes.  Non-frozen
nodes can be either irrelevant or relevant. 
A relevant node influences at
least one other relevant node. 
The relevant nodes completely determine the dynamics and are independent of the
behavior of the irrelevant ones, while the behavior of the irrelevant ones are
completely determined by the relevant ones.  
The set of relevant nodes can be partitioned
into components that are connected subgraphs. Each of these subgraphs is a, so-called,
``relevant component.'' The dynamics of each relevant component is independent of
the others.

The attractor period is the number of synchronous update steps needed until the
same network state occurs again. All possible attractor periods of a network
realization can be deduced by combinatorics involving the possible period
of the dynamics of the relevant components.  In particular, each attractor period is a least
common multiple ($\lcm$) of possible periods for each relevant
component of the realization. For example, if there are two relevant components
with possible periods~$p_1 \in \{2,3,6\}$ and $p_2\in \{1,2\}$, then the possible
attractor periods of the entire network are $P \in \{ 2,3,6,12\}$. 

For critical networks, it is known that almost all relevant components
are simple loops with only the largest component possibly being more complex
\cite{kaufman.drossel:relevant}. 
Using these facts, the dynamics of a network realization can be analyzed as
follows.  In a relevant component that is an ordered loop of $L$ nodes,
every node~$i$ has exactly one input from node~$(i-1)$ for $1<i<L+1$, with
the closing condition $s_{L+1}\equiv s_1$.  In this case, the behavior of
node $i$ is determined by the input it receives from node $(i-1)$. 
\marginpar{(T)}
None of the nodes in the loop can have a Boolean function that gives a
constant output regardless of the state of the previous node in the loop.
Constant outputs would block the loop and immediately lead to a fixed point
attractor.  Thus, two possible coupling functions are left for each node,
either ``copy'' or ``invert''.  

Loops are either ``even'' or ``odd'' depending on the number of
invert-functions they contain. This distinction can be reduced to the
question of whether or not a loop has a single inverting Boolean function,
$f_i=1-s_{i-1}$, because the dynamics of loops with pairs of inverting
functions are equivalent to those of loops without the pairs.  To see this
simply choose two arbitrary nodes with inverting functions and change them
to copy functions, then flip all values of the intermediate nodes. The
number and period of attractors is not changed by this substitution.

Every synchronous update of a loop can be imagined as an incremental
rotation of the whole configuration.  In an even loop, after $N$~updates
the same configuration is reached again.  While in an odd loop, the same
configuration is reached again after $2N$ updates.  However, the periods of
loops with a non-prime-number of nodes can be shorter.
For example, on an even loop, having only copy-functions, consisting of
4~nodes the pattern `0101' has a period of 2.  However, in the arguments
below this possibility is ignored.

Loops with a non-prime-number of nodes can have an even shorter attractor
period.  For example, on an even loop (with only copy-functions) consisting
of 4~nodes the pattern `0101' has a period of 2.  However, such
cases of shorter periods than for prime loops can be ignored as they become
more improbable the more larger components appear.  To justify this,
consider a loop of length $L$.  Let $\mathcal{D}$ be the set of divisors of
$L$.  Shorter attractors have periods that are elements of this set.
Since, as mentioned before, we exclude the two fixed point attractors with
all nodes having the same values, there are $2^L-2$ possible states for a
loop of this length.  How many of these possible states are realized as
part of a shorter period attractor?

Clearly, if $L$ is a prime number, then the cardinality of the set of divisors
is $|\mathcal{D}|=2$. This observable grows extremely slowly with growing
$L$, e.g., $\max_{L\leq 10^4} \left( |\mathcal{D}(L)| \right) = 64$ and only
$\max_{L\leq 10^8} \left( |\mathcal{D}(L)| \right) = 768$. Thus, the
number of divisors of $L$ is much smaller than $L$ itself.
This implies that shorter periods of a loop do not occur very often. 
The probability to have a period smaller than $L$, $P_{<L}$, can be 
estimated as follows.
In principle, the fraction of states which are neither fixed points nor
part of a period of length~$L$ must be calculated. \marginpar{(D)} The
number of those states is given by the length of each shorter period
multiplied with its multiplicity, i.e.{} how often such a period occurs. In
the worst case, the period is $L/2$, which is only possible for even number
of nodes in the loop.  For $|\mathcal{D}| = 3$ (only one additional divisor
beside $1$ and $L$), $P_{<L}$ is maximal, because all states of the period
less than $L$ are united at the only non-trivial divisor value, $L/2$.
Thus, since there are $2^{L/2}$ different patterns of that length, an upper
bound for the fraction of states in an attractor with period shorter than
$L$ is
\begin{eqnarray*}
P_{<L} \leq \frac{2^{L/2}}{2^L-2} \approx 2^{-L/2}
\end{eqnarray*}
Note that the
probability~$P_{<L}$ vanishes for growing $L$.  A similar
argument incorporating the prime number density has been used in
to evaluate a lower bound for the mean
attractor period in $K=1$ networks~\cite{drossel.mihaljev.ea:number}.

\typeout{Results for critical networks} 
The topic of determining the average period $\langle L \rangle$ and average
number $\langle \nu \rangle$ of attractors in critical RBNs has a long history
\cite{kauffman:homeostasis,bastolla:modular,bilke:stability,socolar:scaling}.
\marginpar{(R)}
Only recently it has been shown that both of these quantities increase with
network size $N$ faster than any power law
\cite{samuelsson:superpolynomial,kaufman.mihaljev.ea:scaling}.  Determining the
attractor distribution, however, has received less attention.

In \cite{bhattacharjya.liang:power-law} a numerical algorithm to study the
attractor distribution of Boolean networks was proposed.  Using that
algorithm and studying networks of size up to $N \sim 10^5$, they found
that the attractor periods are power-law distributed for unbiased random
networks with $K=2$, which they interpreted as evidence for the existence
as evidence that these networks are at a critical point at the ``edge of
chaos''.  Other, mostly numerical, studies of biased random networks with
$K=4$ found power-law attractor period distributions on the critical line
at the boundary between the fixed and chaotic phases
\cite{bastolla:modular,bastolla.parisi:numerical,bastolla:closing}.  
%
%
%
It has also been found that the attractor periods are power-law
distributed in the 
self-organized stationary state of Boolean networks whose
set of node functions ${\cal F}$ evolve through a competition  
that punishes majority behavior
\cite{paczuski.bassler.ea:self-organized}.
Because of this, it was concluded that the stationary state is critical.
Generally, however, as was done in these studies and others~\cite{drossel:random}, 
it is possible to 
directly measure the attractor period distributions only for
relatively small networks.
In the following we offer an analytical understanding of the power-law
distribution of attractor periods 
for all kinds of critical networks by explaining how this behavior arises.

\typeout{Calculating the attractor distribution} 
In order to calculate the attractor distribution of a critical network, we
start by looking at the topology of such a network, more precisely at the
distribution of relevant component loops. 
An essential result from previous studies is that the
distribution is Poissonian, and that a loop consisting of
$L$~nodes appears with probability $p_{L}=L^{-1}$ when $L$ is smaller than a
cutoff, $L < L_{\rm max}$.  The cut-off length depends on the size of
the network, $L_{\rm max} \sim \sqrt{N}$ \cite{drossel:number}. 

The period of the attractor is determined by the $\lcm$ of all size
relevant components.  As discussed above, it is known that, in the large
network limit, only the largest relevant component of a critical network
has a finite probability of being more complex than a simple
connection-loop \cite{kaufman.drossel:relevant}. 

We now make the approximation that \emph{all} relevant components are simple loops.
This assumption is valid because the complex components
can be constructed from additional links in loops or by interconnecting
multiple loops.  Depending on the network topology and 
on the Boolean functions at the nodes with more than one relevant input, the
period of the complex component can be either smaller or larger than the
length of the loops which are used to construct it
\cite{kaufman.drossel:relevant}. Shorter period components will be included by our
random procedure by just picking smaller loop lengths. For larger periods,
complex relevant components deliver a contribution which can play a role
only when their period is comparable with $L_{\rm max}$.  Thus, the
validity of the assumption that all relevant components are simple loops
improves as $L_{\rm max}$ increases.

We also assume that the period of the attractor corresponding
to a given loop is equal to the length of the loop.
This assumption applies even to loops with odd length that actually
have a period that is twice as long.
However, the factor of 2 that occurs for odd length loops, produces only a
small correction to our results and does not change our principal 
conclusions.

Thus, we conclude, that the distribution of attractor periods decays as a
power law with exponent $-1$, $p(L) \sim 1/L$. As discussed above, this
result is consistent with previous numerical studies.

\begin{figure}[b!]
\begin{center}
\includegraphics[width=0.95\columnwidth]{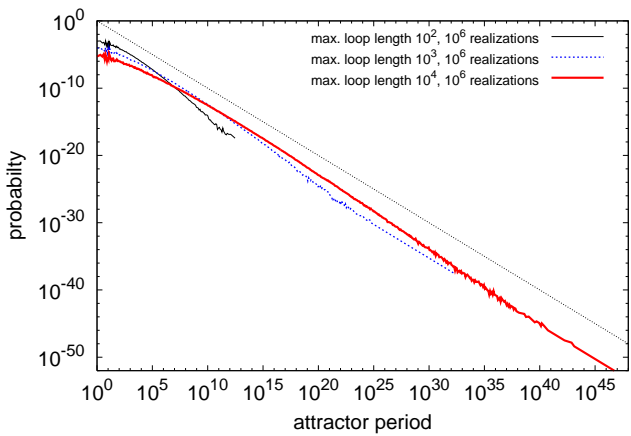}
\end{center}
\caption[Distribution of attractor lengths.]{Distribution of attractor
periods~$L$. These probability distributions are found by generating relevant
components and deducing the corresponding attractor period for
$10^6$~realizations. The known loop-distribution is used, up to a cutoff value.
The figure shows different cut-offs of $10^2$ (blue slashed line), $10^3$ (red
crosses) and $10^4$ (solid black line).  It is assumed that each loop
contributes only to one attractor that in turn has an attractor period
comparable to its number of nodes.  The dotted line corresponds to a power-law
with exponent $-1$.  The fluctuations at small attractor periods are due to the
number of divisors integers have, see text for details.} \label{FigSampling}
\end{figure}

The validity of the above arguments can be explored
using numerical sampling.
Take $s$~different samples, each of which corresponds to a single attractor
of a network realization. 
Note that a network realization might have more than one possible
attractor, but we take only one per realization.
Then, the procedure described above is implemented: Take a set of loops of
length $L$, each of which occur with probability $1/L$, and calculate the $\lcm$. 
Using this method we end up with an histogram of attractor period, as shown in
Fig.~\ref{FigSampling}.

In order to display the histogram data in a logarithmic representation, the
data is binned, i.e., attractor periods within a certain range are put into one
bin. The width of the bins grows with a binning factor, a given bin has $b$
times the size of its neighboring bin on the  left. The results are shown
with a binning factor of $b=1.2$ for $L>14$, for small $L$ we just used the
period of the attractor itself. 
By this choice we guarantee that each bin contains at least one possible attractor
period. Binning is used as soon the bins may contain more than one
possible attractor period, i.e., for attractors with more 14 states.
The histogram is normalized such that the total probability is one. 
As expected, as the network size, or equivalently the maximum loop length, grows
the distribution approaches a power law with exponent $-1$.
Note that with this new sampling method huge system
sizes can be studied. 
The free parameter in our method is the cutoff~$L_{\rm max}$ which is a
function of the system size. 
A further simplification is to take just the product of the individual loop
periods instead of the $\lcm$.

\typeout{> Divisibility effects.} 

\begin{figure}[h!]
\begin{center}
{\includegraphics[width=0.95\columnwidth]{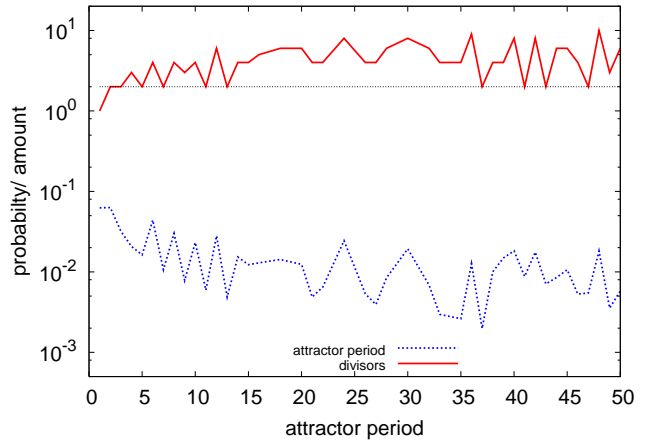} }
\end{center}\vspace{-0.5cm}
\caption{Probability of attractors to have period $L$ (blue dashed curve)
and the number of divisors of $L$ (red solid curve).  Each time the values
in the upper curve have the value 2 (dotted horizontal line), $L$ is a
prime number.} \label{FigDivisors}
\end{figure}

Consider what happens when the attractor period is short. 
\marginpar{(C)}
Figure~\ref{FigDivisors} shows the result
using the $\lcm$ method. 
It also shows the number of divisors
of each integer.
Binning is not used in this figure.
The distribution has spikes at particular periods, which are also
apparent in
Fig.~\ref{FigSampling}. These spikes occur when the number of
divisors is small.  The
histogram for the distribution of attractor periods in the self-organized
steady state of an evolutionary
Boolean network model, shown in Fig.~{2} of
\cite{paczuski.bassler.ea:self-organized}, exhibits a very similar
spike-pattern.  Note that in that paper, binning was used for smaller
attractor periods, thus some peaks are averaged away.

As explained above, a given attractor 
period can be approximated by taking the least common multiple of 
some loop lengths (as proxy for the loop period) and we also
know the probability for each loop length. 
If \emph{all} loops up to certain length~$x$ contribute to the attractor,
number theory provides a formula for that. 
Using the prime number theorem, and an inequality proved by Nair
\cite{tenenbaum:introduction}, the least
common multiples of the first $x$ positive integers with $x=\{1,2,\ldots
\}$ obey
\begin{eqnarray}
s(x) &\equiv& \lcm(1,2,\ldots,x) \geq \exp\left( x \left(1+\mathcal{O}(1)\right)\right)
\label{EqNumberTheory}
\end{eqnarray}
The series~$s(x)$ starts with $1, 2, 6, 12, 60, 420, \ldots$ \cite{sloane:on-line} and
represents all possible attractor periods.
For large $x$, the
probability for an attractor period constructed of \emph{all} possible periods
is negligible because the $e^{x}$ in Eq.~(\ref{EqNumberTheory}) appears
in the denominator of the probabilities for the overall attractor period.
For smaller $x$, of say $x<10^3$, this approximation does
not hold yet, but then the approximation $p(L) \sim 1/L$ does hold.

Only a full state space enumeration of the dynamics allows one to obtain the
exact attractor distribution. However, this is only possible for small
system sizes.  This is true even if an intelligent pruning algorithm is used that
disregards irrelevant nodes and simulates only the dynamics of the relevant nodes
for a given realization.

For this reason, previous studies of the attractor period distribution have
relied on sampling. However, sampling has potential problems \cite{
berdahl.shreim.ea:random}. One problem that can occur when generating
attractor-statistics by sampling of various network realizations is
undersampling. Undersampling occurs when simulating without any prior
knowledge about the structure of the state space. If nothing about the
relevant components is known, then one has to determine to which attractor
each initial condition converges to.  In order to do this, one has to
determine the successor for each state, meaning $2^{N}$ updates.  Because
of this restriction, it is only possible to sample only a set of initial
configurations that correspond to a negligible fraction of the state space
for large system sizes.  Another known problem that occurs with sampling is
that the frequency with which attractors are found depends on the size of
their basin of attraction, see e.g. Ref.
\cite{samuelsson:superpolynomial}.

Our new method has neither the problem of undersampling nor of being biased by
the basin sizes, and allows us to effectively study very large
networks. It constructs the overall attractor of a
network realization by taking the period of the relevant components the
realization is constructed of.  Our analytic arguments explain the 
numerical evidence found by
others that attractor periods in large critical Boolean networks are
power-law distributed. Thus, critical Boolean networks exhibit
scaling also in the attractor period distribution, a property that until now has not
been analytically shown.

\begin{acknowledgments}
The authors gratefully acknowledge useful discussions with Barbara Drossel.
The work of FG was supported by the German Research Foundation (Deutsche
Forschungsgemeinschaft, DFG) under contract No.\ Dr300/4.
The work of KEB was supported by the NSF through grant No.\ DMR-0908286 
and by the Texas Advanced Research Program through grant No.\ 95921.
\end{acknowledgments}

\bibliography{fgAld_Bib.bib}

\begin{thebibliography}{33}
\expandafter\ifx\csname natexlab\endcsname\relax\def\natexlab#1{#1}\fi
\expandafter\ifx\csname bibnamefont\endcsname\relax
  \def\bibnamefont#1{#1}\fi
\expandafter\ifx\csname bibfnamefont\endcsname\relax
  \def\bibfnamefont#1{#1}\fi
\expandafter\ifx\csname citenamefont\endcsname\relax
  \def\citenamefont#1{#1}\fi
\expandafter\ifx\csname url\endcsname\relax
  \def\url#1{\texttt{#1}}\fi
\expandafter\ifx\csname urlprefix\endcsname\relax\def\urlprefix{URL }\fi
\providecommand{\bibinfo}[2]{#2}
\providecommand{\eprint}[2][]{\url{#2}}

\bibitem[{\citenamefont{Albert and Barab\'{a}si}(2002)}]{albert:statistical}
\bibinfo{author}{\bibfnamefont{R.}~\bibnamefont{Albert}} \bibnamefont{and}
  \bibinfo{author}{\bibfnamefont{A.-L.} \bibnamefont{Barab\'{a}si}},
  \bibinfo{journal}{Rev. Mod. Phys.} \textbf{\bibinfo{volume}{74}},
  \bibinfo{pages}{47} (\bibinfo{year}{2002}).

\bibitem[{\citenamefont{Aldana-Gonzalez
  et~al.}(2003)\citenamefont{Aldana-Gonzalez, Coppersmith, and
  Kadanoff}}]{aldana-gonzalez:boolean}
\bibinfo{author}{\bibfnamefont{M.}~\bibnamefont{Aldana-Gonzalez}},
  \bibinfo{author}{\bibfnamefont{S.}~\bibnamefont{Coppersmith}},
  \bibnamefont{and} \bibinfo{author}{\bibfnamefont{L.~P.}
  \bibnamefont{Kadanoff}}, \bibinfo{journal}{Perspectives and Problems in
  Nonlinear Science} pp. \bibinfo{pages}{23--89} (\bibinfo{year}{2003}).

\bibitem[{\citenamefont{Drossel}(2008)}]{drossel:random}
\bibinfo{author}{\bibfnamefont{B.}~\bibnamefont{Drossel}}, in
  \emph{\bibinfo{booktitle}{Reviews of Nonlinear Dynamics and Complexity}},
  edited by \bibinfo{editor}{\bibfnamefont{H.-G.} \bibnamefont{Schuster}}
  (\bibinfo{publisher}{Wiley}, \bibinfo{year}{2008}), vol.~\bibinfo{volume}{1},
  pp. \bibinfo{pages}{69--110}, ISBN \bibinfo{isbn}{978-3-527-40729-3}.

\bibitem[{\citenamefont{Li et~al.}(2004)\citenamefont{Li, Long, Lu, Ouyang, and
  Tang}}]{li.long.ea:yeast}
\bibinfo{author}{\bibfnamefont{F.}~\bibnamefont{Li}},
  \bibinfo{author}{\bibfnamefont{T.}~\bibnamefont{Long}},
  \bibinfo{author}{\bibfnamefont{Y.}~\bibnamefont{Lu}},
  \bibinfo{author}{\bibfnamefont{Q.}~\bibnamefont{Ouyang}}, \bibnamefont{and}
  \bibinfo{author}{\bibfnamefont{C.}~\bibnamefont{Tang}},
  \textbf{\bibinfo{volume}{101}}, \bibinfo{pages}{4781} (\bibinfo{year}{2004}).

\bibitem[{\citenamefont{Albert and Othmer}(2003)}]{albert.othmer:topology}
\bibinfo{author}{\bibfnamefont{R.}~\bibnamefont{Albert}} \bibnamefont{and}
  \bibinfo{author}{\bibfnamefont{H.~G.} \bibnamefont{Othmer}},
  \bibinfo{journal}{J. Theo. Bio.} \textbf{\bibinfo{volume}{223}},
  \bibinfo{pages}{1} (\bibinfo{year}{2003}).

\bibitem[{\citenamefont{Davidich and
  Bornholdt}(2008)}]{davidich.bornholdt:boolean}
\bibinfo{author}{\bibfnamefont{M.~I.} \bibnamefont{Davidich}} \bibnamefont{and}
  \bibinfo{author}{\bibfnamefont{S.}~\bibnamefont{Bornholdt}},
  \bibinfo{journal}{PLoS ONE} \textbf{\bibinfo{volume}{3}},
  \bibinfo{pages}{1672} (\bibinfo{year}{2008}).

\bibitem[{\citenamefont{Bornholdt}(2005)}]{bornholdt:less}
\bibinfo{author}{\bibfnamefont{S.}~\bibnamefont{Bornholdt}},
  \bibinfo{journal}{Science} \textbf{\bibinfo{volume}{310}},
  \bibinfo{pages}{449} (\bibinfo{year}{2005}).

\bibitem[{\citenamefont{Samuelsson and
  Troein}(2003)}]{samuelsson:superpolynomial}
\bibinfo{author}{\bibfnamefont{B.}~\bibnamefont{Samuelsson}} \bibnamefont{and}
  \bibinfo{author}{\bibfnamefont{C.}~\bibnamefont{Troein}},
  \bibinfo{journal}{Phys. Rev. Lett.} \textbf{\bibinfo{volume}{90}},
  \bibinfo{pages}{098701} (\bibinfo{year}{2003}).

\bibitem[{\citenamefont{Drossel}(2005)}]{drossel:number}
\bibinfo{author}{\bibfnamefont{B.}~\bibnamefont{Drossel}},
  \bibinfo{journal}{Phys. Rev. E} \textbf{\bibinfo{volume}{72}},
  \bibinfo{pages}{016110} (\bibinfo{year}{2005}).

\bibitem[{\citenamefont{Mihaljev and Drossel}(2006)}]{mihaljev.drossel:scaling}
\bibinfo{author}{\bibfnamefont{T.}~\bibnamefont{Mihaljev}} \bibnamefont{and}
  \bibinfo{author}{\bibfnamefont{B.}~\bibnamefont{Drossel}},
  \bibinfo{journal}{Phys. Rev. E} \textbf{\bibinfo{volume}{74}},
  \bibinfo{pages}{046101} (\bibinfo{year}{2006}).

\bibitem[{\citenamefont{Drossel et~al.}(2005)\citenamefont{Drossel, Mihaljev,
  and Greil}}]{drossel.mihaljev.ea:number}
\bibinfo{author}{\bibfnamefont{B.}~\bibnamefont{Drossel}},
  \bibinfo{author}{\bibfnamefont{T.}~\bibnamefont{Mihaljev}}, \bibnamefont{and}
  \bibinfo{author}{\bibfnamefont{F.}~\bibnamefont{Greil}},
  \bibinfo{journal}{Phys. Rev. Lett.} \textbf{\bibinfo{volume}{94}},
  \bibinfo{pages}{088701} (\bibinfo{year}{2005}).

\bibitem[{\citenamefont{Paczuski et~al.}(2000)\citenamefont{Paczuski, Bassler,
  and Corral}}]{paczuski.bassler.ea:self-organized}
\bibinfo{author}{\bibfnamefont{M.}~\bibnamefont{Paczuski}},
  \bibinfo{author}{\bibfnamefont{K.~E.} \bibnamefont{Bassler}},
  \bibnamefont{and} \bibinfo{author}{\bibfnamefont{A.}~\bibnamefont{Corral}},
  \bibinfo{journal}{Phys. Rev. Lett.} \textbf{\bibinfo{volume}{84}},
  \bibinfo{pages}{3185} (\bibinfo{year}{2000}).

\bibitem[{\citenamefont{Bassler et~al.}(2004)\citenamefont{Bassler, Lee, and
  Lee}}]{bassler.lee.ea:evolution}
\bibinfo{author}{\bibfnamefont{K.~E.} \bibnamefont{Bassler}},
  \bibinfo{author}{\bibfnamefont{C.}~\bibnamefont{Lee}}, \bibnamefont{and}
  \bibinfo{author}{\bibfnamefont{Y.}~\bibnamefont{Lee}},
  \bibinfo{journal}{Phys. Rev. Lett.} \textbf{\bibinfo{volume}{93}},
  \bibinfo{pages}{038101} (\bibinfo{year}{2004}).

\bibitem[{\citenamefont{Liu and Bassler}(2007)}]{liu.bassler:finite}
\bibinfo{author}{\bibfnamefont{M.}~\bibnamefont{Liu}} \bibnamefont{and}
  \bibinfo{author}{\bibfnamefont{K.~E.} \bibnamefont{Bassler}},
  \bibinfo{journal}{{\tt arXiv:0711.2314}}  (\bibinfo{year}{2007}).

\bibitem[{\citenamefont{Bornholdt and
  Rohlf}(2000)}]{bornholdt.rohlf:topological}
\bibinfo{author}{\bibfnamefont{S.}~\bibnamefont{Bornholdt}} \bibnamefont{and}
  \bibinfo{author}{\bibfnamefont{T.}~\bibnamefont{Rohlf}},
  \bibinfo{journal}{Phys. Rev. Lett.} \textbf{\bibinfo{volume}{84}},
  \bibinfo{pages}{6114} (\bibinfo{year}{2000}).

\bibitem[{\citenamefont{Liu and Bassler}(2006)}]{liu.bassler:emergent}
\bibinfo{author}{\bibfnamefont{M.}~\bibnamefont{Liu}} \bibnamefont{and}
  \bibinfo{author}{\bibfnamefont{K.~E.} \bibnamefont{Bassler}},
  \bibinfo{journal}{Phys. Rev. E} \textbf{\bibinfo{volume}{74}},
  \bibinfo{pages}{041910} (\bibinfo{year}{2006}).

\bibitem[{\citenamefont{Gross and Blasius}(2008)}]{gross.blasius:adaptive}
\bibinfo{author}{\bibfnamefont{T.}~\bibnamefont{Gross}} \bibnamefont{and}
  \bibinfo{author}{\bibfnamefont{B.}~\bibnamefont{Blasius}},
  \bibinfo{journal}{J. R. Soc. Interface} \textbf{\bibinfo{volume}{5}},
  \bibinfo{pages}{259} (\bibinfo{year}{2008}).

\bibitem[{\citenamefont{Luque et~al.}(2001)\citenamefont{Luque, Ballesteros,
  and Muro}}]{luque.ballesteros.ea:self-organized}
\bibinfo{author}{\bibfnamefont{B.}~\bibnamefont{Luque}},
  \bibinfo{author}{\bibfnamefont{F.~J.} \bibnamefont{Ballesteros}},
  \bibnamefont{and} \bibinfo{author}{\bibfnamefont{E.~M.} \bibnamefont{Muro}},
  \bibinfo{journal}{Phys. Rev. E} \textbf{\bibinfo{volume}{63}},
  \bibinfo{pages}{051913} (\bibinfo{year}{2001}).

\bibitem[{\citenamefont{Reichhardt and
  Bassler}(2007)}]{reichhardt.bassler:canalization}
\bibinfo{author}{\bibfnamefont{C.~J.~O.} \bibnamefont{Reichhardt}}
  \bibnamefont{and} \bibinfo{author}{\bibfnamefont{K.~E.}
  \bibnamefont{Bassler}}, \bibinfo{journal}{J. Phys. A}
  \textbf{\bibinfo{volume}{40}}, \bibinfo{pages}{4339} (\bibinfo{year}{2007}).

\bibitem[{\citenamefont{Derrida and Pomeau}(1986)}]{derrida:random}
\bibinfo{author}{\bibfnamefont{B.}~\bibnamefont{Derrida}} \bibnamefont{and}
  \bibinfo{author}{\bibfnamefont{Y.}~\bibnamefont{Pomeau}},
  \bibinfo{journal}{Europhys. Lett.} \textbf{\bibinfo{volume}{1}},
  \bibinfo{pages}{45} (\bibinfo{year}{1986}).

\bibitem[{\citenamefont{Shmulevich and
  Kauffman}(2004)}]{shmulevich.kauffman:activities}
\bibinfo{author}{\bibfnamefont{I.}~\bibnamefont{Shmulevich}} \bibnamefont{and}
  \bibinfo{author}{\bibfnamefont{S.~A.} \bibnamefont{Kauffman}},
  \bibinfo{journal}{Phys. Rev. Lett.} \textbf{\bibinfo{volume}{93}},
  \bibinfo{pages}{048701} (\bibinfo{year}{2004}).

\bibitem[{\citenamefont{Bastolla and Parisi}(1998)}]{bastolla:modular}
\bibinfo{author}{\bibfnamefont{U.}~\bibnamefont{Bastolla}} \bibnamefont{and}
  \bibinfo{author}{\bibfnamefont{G.}~\bibnamefont{Parisi}},
  \bibinfo{journal}{Physica D} \textbf{\bibinfo{volume}{115}},
  \bibinfo{pages}{219} (\bibinfo{year}{1998}).

\bibitem[{\citenamefont{Kaufman and Drossel}(2006)}]{kaufman.drossel:relevant}
\bibinfo{author}{\bibfnamefont{V.}~\bibnamefont{Kaufman}} \bibnamefont{and}
  \bibinfo{author}{\bibfnamefont{B.}~\bibnamefont{Drossel}},
  \bibinfo{journal}{New. J. Phys.} \textbf{\bibinfo{volume}{9}},
  \bibinfo{pages}{228} (\bibinfo{year}{2006}).

\bibitem[{\citenamefont{Kauffman}(1969)}]{kauffman:homeostasis}
\bibinfo{author}{\bibfnamefont{S.}~\bibnamefont{Kauffman}},
  \bibinfo{journal}{Nature} \textbf{\bibinfo{volume}{224}},
  \bibinfo{pages}{177} (\bibinfo{year}{1969}).

\bibitem[{\citenamefont{Bilke and Sjunnesson}(2002)}]{bilke:stability}
\bibinfo{author}{\bibfnamefont{S.}~\bibnamefont{Bilke}} \bibnamefont{and}
  \bibinfo{author}{\bibfnamefont{F.}~\bibnamefont{Sjunnesson}},
  \bibinfo{journal}{Phys. Rev. E} \textbf{\bibinfo{volume}{65}},
  \bibinfo{pages}{016129} (\bibinfo{year}{2002}).

\bibitem[{\citenamefont{Socolar and Kauffman}(2003)}]{socolar:scaling}
\bibinfo{author}{\bibfnamefont{J.~E.~S.} \bibnamefont{Socolar}}
  \bibnamefont{and} \bibinfo{author}{\bibfnamefont{S.~A.}
  \bibnamefont{Kauffman}}, \bibinfo{journal}{Phys. Rev. Lett.}
  \textbf{\bibinfo{volume}{90}}, \bibinfo{pages}{068702}
  (\bibinfo{year}{2003}).

\bibitem[{\citenamefont{Kaufman et~al.}(2005)\citenamefont{Kaufman, Mihaljev,
  and Drossel}}]{kaufman.mihaljev.ea:scaling}
\bibinfo{author}{\bibfnamefont{V.}~\bibnamefont{Kaufman}},
  \bibinfo{author}{\bibfnamefont{T.}~\bibnamefont{Mihaljev}}, \bibnamefont{and}
  \bibinfo{author}{\bibfnamefont{B.}~\bibnamefont{Drossel}},
  \bibinfo{journal}{Phys. Rev. E} \textbf{\bibinfo{volume}{72}},
  \bibinfo{pages}{046124} (\bibinfo{year}{2005}).

\bibitem[{\citenamefont{Bhattacharjya and
  Liang}(1996)}]{bhattacharjya.liang:power-law}
\bibinfo{author}{\bibfnamefont{A.}~\bibnamefont{Bhattacharjya}}
  \bibnamefont{and} \bibinfo{author}{\bibfnamefont{S.}~\bibnamefont{Liang}},
  \bibinfo{journal}{Phys. Rev. Lett.} \textbf{\bibinfo{volume}{77}},
  \bibinfo{pages}{1644} (\bibinfo{year}{1996}).

\bibitem[{\citenamefont{Bastolla and Parisi}(1997)}]{bastolla.parisi:numerical}
\bibinfo{author}{\bibfnamefont{U.}~\bibnamefont{Bastolla}} \bibnamefont{and}
  \bibinfo{author}{\bibfnamefont{G.}~\bibnamefont{Parisi}},
  \bibinfo{journal}{J. Theo. Bio.} \textbf{\bibinfo{volume}{187}},
  \bibinfo{pages}{117} (\bibinfo{year}{1997}).

\bibitem[{\citenamefont{Bastolla and Parisi}(1996)}]{bastolla:closing}
\bibinfo{author}{\bibfnamefont{U.}~\bibnamefont{Bastolla}} \bibnamefont{and}
  \bibinfo{author}{\bibfnamefont{G.}~\bibnamefont{Parisi}},
  \bibinfo{journal}{Physica D} \textbf{\bibinfo{volume}{98}},
  \bibinfo{pages}{1} (\bibinfo{year}{1996}).

\bibitem[{\citenamefont{Tenenbaum}(1995)}]{tenenbaum:introduction}
\bibinfo{author}{\bibfnamefont{G.}~\bibnamefont{Tenenbaum}},
  \emph{\bibinfo{title}{Introduction to Analytic and Probabilistic Number
  Theory}} (\bibinfo{publisher}{Cambridge University Press},
  \bibinfo{year}{1995}).

\bibitem[{\citenamefont{Sloane and Plouffe}(1995)}]{sloane:on-line}
\bibinfo{author}{\bibfnamefont{N.~J.~A.} \bibnamefont{Sloane}}
  \bibnamefont{and} \bibinfo{author}{\bibfnamefont{S.}~\bibnamefont{Plouffe}},
  \emph{\bibinfo{title}{On-line encyclopedia of integer sequences}}
  (\bibinfo{year}{1995}), \bibinfo{note}{{\tt A003418}},
  \urlprefix\url{www.research.att.com/~njas/sequences/}.

\bibitem[{\citenamefont{Berdahl et~al.}(2009)\citenamefont{Berdahl, Shreim,
  Sood, Paczuski, and Davidsen}}]{berdahl.shreim.ea:random}
\bibinfo{author}{\bibfnamefont{A.}~\bibnamefont{Berdahl}},
  \bibinfo{author}{\bibfnamefont{A.}~\bibnamefont{Shreim}},
  \bibinfo{author}{\bibfnamefont{V.}~\bibnamefont{Sood}},
  \bibinfo{author}{\bibfnamefont{M.}~\bibnamefont{Paczuski}}, \bibnamefont{and}
  \bibinfo{author}{\bibfnamefont{J.}~\bibnamefont{Davidsen}},
  \bibinfo{journal}{New. J. Phys.} \textbf{\bibinfo{volume}{11}},
  \bibinfo{pages}{043024} (\bibinfo{year}{2009}).

\end{thebibliography}

\end{document}